**Title:** Extended Phase Graph formalism for systems with Magnetization Transfer and Chemical Exchange

**Authors:** Shaihan J Malik, Rui PAG Teixeira, Joseph V Hajnal

**Correspondence to:**

Dr. Shaihan Malik
Division of Imaging Sciences and Biomedical Engineering
King's College London
1$^{st}$ floor South Wing
St. Thomas' Hospital
London SE1 7EH

Word count (main body) 5,710

Running Title:   EPG-X: Extended Phase Graphs for exchanging systems

**Abstract**

**Purpose:** An Extended Phase Graph framework for modelling systems with exchange or magnetization transfer (MT) is proposed.

**Theory:** The proposed framework (named EPG-X) models coupled two-compartment systems by describing each compartment with separate phase graphs that exchange during evolution periods. There are two variants: EPG-X(BM) for systems governed by the Bloch-McConnell equations; and EPG-X(MT) for the pulsed MT formalism. For the MT case the 'bound' protons have no transverse components so their phase graph consists only longitudinal states.

**Methods:** EPG-X was used to model steady-state gradient echo imaging, MT effects in multislice Turbo Spin Echo imaging, multiecho CPMG for multicomponent $T_2$ relaxometry and transient variable flip angle gradient echo imaging of the type used for MR Fingerprinting. Experimental data were also collected for the final case.

**Results:** Steady-state predictions from EPG-X closely match directly derived steady-state solutions which differ substantially from classic 'single pool' EPG predictions. EPG-X(MT) predicts similar MT related levels of signal attenuation in white matter as have been reported elsewhere in the literature. Modelling of CPMG echo trains with EPG-X(BM) suggests that exchange processes can lead to an underestimate of the fraction of short $T_2$ species. Modelling of transient gradient echo sequences with EPG-X(MT) suggests that measureable MT effects result from variable saturation of bound protons, particularly after inversion pulses.

**Conclusions:** EPG-X can be used for modelling of the transient signal response of systems exhibiting chemical exchange or MT. This may be particularly beneficial for relaxometry approaches that rely on characterising transient rather than steady-state sequences.

**Word count: 249**

**Introduction**

The Extended Phase Graph (EPG) algorithm (1–3) is a commonly used tool for simulating signals obtained from MRI pulse sequences including multiple RF and gradient pulses both qualitatively and quantitatively. It has been used for a diverse and growing range of applications including characterization of RF spoiling in gradient echo sequences (4,5), analysis of echo amplitudes in turbo spin echo (TSE) sequences (6–9), parallel transmission sequence design (10,11), diffusion effects (12), and characterizing signal evolution in sequences used for relaxometry (13–16).

The EPG method is a Fourier approach to solving the Bloch equations, and therefore assumes that tissues are characterised by a single set of relaxation parameters. It is recognised though, that a single compartment approach fails to fully characterise complex biological tissues in many circumstances. Instead coupled multi-compartment models have been proposed. The Bloch-McConnell (BM) equations (17) are a general form for describing systems that are coupled via a general chemical exchange process, with a modification further to describe magnetization transfer (18,19). MT effects in particular have been shown to be strong determinants of observed signals in human tissue – for example ref (20) shows on-resonance MT effects are expected to change the signal from balanced SSFP sequences in brain by approximately 30% and muscle by 50%.

The EPG formalism provides a computationally efficient method for modelling of MR sequences that also gives intuitive insight into signal formation, by isolating different pathways that can lead to echo formation. Currently it is not possible to use EPGs to model multi-compartment systems with exchange (different compartments in non-exchanging systems can simply be modelled separately and then averaged). Hence this work seeks to extend the EPG formalism to model such systems. This is increasingly relevant to emerging transient phase relaxometry approaches such as Magnetic Resonance Fingerprinting (MRF) (21) which require numerical simulation, as opposed to more traditional steady-state methods for which analytic or closed form solutions are often available (for example the DESPOT or multi-component DESPOT methods(22)).

We first outline the proposed extensions to the EPG method, and then test them against known steady-state solutions for gradient echo imaging (used by the multi-component DESPOT method (22)). They are then used to explore some test cases to illustrate the impact of the new approach. Test cases include multislice vs single slice TSE, and two different relaxometry methods - multiecho CPMG data for multicomponent $T_2$ estimation, and gradient echo imaging with modulated flip angles (similar to MRF). In the latter case experimental data were also collected.

## Theory

After a sequence of multiple RF and gradient pulses the magnetization generally forms a complex distribution at the sub-voxel level; the EPG algorithm is a Fourier domain approach to characterising this distribution. In this work, we will limit ourselves to sequences with equidistant timing and an unbalanced gradient that does not change direction – in this case the sub-voxel magnetization distribution may be defined in terms of the gradient induced phase $\psi$ during some fixed time period $\Delta t$. An idealised voxel may be defined by the interval $\psi \in [-\pi, \pi]$ with uniform density of magnetization in this range. Isochromat based simulations model a sequence by sampling this range directly, with the predicted signal being the sum over the ensemble. In the EPG representation the magnetization is represented by configuration states $\tilde{F}_n$ and $\tilde{Z}_n$ which, using the notation from the introductory review from Weigel (3), may be defined as follows:

$$M_+(\psi) = M_x(\psi) + iM_y(\psi) = \sum_{n=-\infty}^{\infty} \tilde{F}_n e^{in\psi}$$

$$M_-(\psi) = M_x(\psi) - iM_y(\psi) = \sum_{n=-\infty}^{\infty} \tilde{F}_n^* e^{in\psi}$$

$$M_z(\psi) = \sum_{n=-\infty}^{\infty} \tilde{Z}_n e^{in\psi}$$

. [1]

In this case the idealised voxel means that the received signal at a given time is simply given by the value of $\tilde{F}_0$ with no transverse states with n≠0 contributing any signal.

## Bloch McConnell Equations for chemical exchange

Consider a system with two compartments arbitrarily labelled *a* and *b* with thermal equilibrium magnetizations $M_0^a = (1-f)M_0$ and $M_0^b = fM_0$ respectively, where $M_0$ is the total magnetization and *f* is the fraction in compartment *b* which is conventionally assumed to be smaller. When not at thermal equilibrium the components of the magnetization may be written as vector $\boldsymbol{M} = \begin{bmatrix} M_x^a & M_y^a & M_z^a & M_x^b & M_y^b & M_z^b \end{bmatrix}^T$ whose time evolution is governed by $\dot{\boldsymbol{M}} = \boldsymbol{A}\boldsymbol{M} + \boldsymbol{C}$. Following the notation used by Zaiss (23) the terms can be written (in the rotating frame) as follows:

$$\boldsymbol{A} = \begin{bmatrix} \boldsymbol{L}_a - \boldsymbol{K}_a & +\boldsymbol{K}_b \\ +\boldsymbol{K}_a & \boldsymbol{L}_b - \boldsymbol{K}_b \end{bmatrix} \quad \boldsymbol{K}_j = \begin{bmatrix} k_j & 0 & 0 \\ 0 & k_j & 0 \\ 0 & 0 & k_j \end{bmatrix} \quad \boldsymbol{L}_j = \begin{bmatrix} -R_{2,j} & +\omega_z & -\omega_y \\ -\omega_z & -R_{2,j} & \omega_x \\ \omega_y & -\omega_x & -R_{1,j} \end{bmatrix}$$

$$C = \begin{bmatrix} 0 & 0 & R_{1,a}M_0^a & 0 & 0 & R_{1,b} & M_0^b \end{bmatrix}^T \qquad j \in \{a,b\} \qquad [2]$$

In these expressions, $k_a$ is the exchange rate from *a* to *b*, which is related to the reverse exchange rate $k_b$ via $k_a M_0^a = k_b M_0^b$ to preserve balance at thermal equilibrium.

The BM equations imply that during a pulse sequence the overall state of the magnetization must now be characterised separately for the two compartments, and hence the EPG representation must be extended to also reflect this; i.e. we must have 'states' that correspond to magnetization from *a* and *b*, $\left[ \tilde{F}_n^a \; \tilde{F}_{-n}^{*\,a} \; \tilde{Z}_n^a \; \tilde{F}_n^b \; \tilde{F}_{-n}^{*\,b} \; \tilde{Z}_n^b \right]^T$.

**Evolution in absence of RF**

The EPG formalism solves the Bloch equations by treating RF pulses as instantaneous - RF pulses are thus treated separately from the evolution periods in which gradients and relaxation effects are considered. We follow the same approach here, with the addition of exchange terms during the evolution periods where $\omega_x = \omega_y = 0$.

There are essentially two steps necessary to move from a simple 'isochromat' picture of the magnetization vector $[M_x \; M_y \; M_z]$ to the EPG representation: i) change basis to $[M_+ \; M_- \; M_z]$ and ii) apply Fourier Transform to move to $[\tilde{F}_n \; \tilde{F}_{-n}^* \; \tilde{Z}_n]$. In deriving the standard EPG formalism (3) the change of basis is achieved by applying similarity transform **σ**

$$\sigma = \begin{bmatrix} 1 & i & 0 \\ 1 & -i & 0 \\ 0 & 0 & 1 \end{bmatrix} \qquad [3]$$

which we now extend by creating block diagonal matrix

$$S = \begin{bmatrix} \sigma & 0 \\ 0 & \sigma \end{bmatrix} \qquad [4]$$

In this basis we represent the magnetization as $M = \begin{bmatrix} M_+^a & M_-^a & M_z^a & M_+^b & M_-^b & M_z^b \end{bmatrix}^T$ and the system matrix becomes:

$$A^\dagger = SAS^{-1} = \begin{bmatrix} L_a^\dagger - K_a & +K_b \\ +K_a & L_b^\dagger - K_b \end{bmatrix} \qquad [5]$$

where the '†' superscripts indicate the change of basis. Matrices **K**$_{a,b}$ are unchanged, however $L_j^\dagger$ is written as

$$L_j^\dagger = \begin{bmatrix} -R_{2,j} - i\omega_z & 0 & 0 \\ 0 & -R_{2,j} + i\omega_z & 0 \\ 0 & 0 & -R_{1,j} \end{bmatrix}$$

$$= \begin{bmatrix} -R_{2,j} & 0 & 0 \\ 0 & -R_{2,j} & 0 \\ 0 & 0 & -R_{1,j} \end{bmatrix} + \begin{bmatrix} -i\omega_z & 0 & 0 \\ 0 & +i\omega_z & 0 \\ 0 & 0 & 0 \end{bmatrix} = E_j + W \qquad [6]$$

where **E$_j$** is the familiar relaxation matrix from the standard EPG algorithm and **W** accounts for gradient induced dephasing (and off resonance). Hence the full evolution matrix is written:

$$A^\dagger = \begin{bmatrix} E_a - K_a & +K_b \\ +K_a & E_b - K_b \end{bmatrix} + \begin{bmatrix} W & 0 \\ 0 & W \end{bmatrix} = \Lambda + \Omega \qquad [7]$$

Finally, we note that the transverse and longitudinal components do not interact (i.e. $M_z^a$ and $M_z^b$ couple only to each other and not the transverse components) so the matrices can be separated to obtain:

$$\dot{M}_T = (\Lambda_T + \Omega_T) M_T$$

$$\dot{M}_L = \Lambda_L M_L + C_L$$

$$\qquad [8]$$

Where $\boldsymbol{M_T} = [M_+^a \ M_-^a \ M_+^b \ M_-^b]^T$ and $\boldsymbol{M_L} = [M_z^a \ M_z^b]^T$ correspond to transverse and longitudinal components respectively, and the other matrices now written out in full are:

$$\Lambda_T = \begin{bmatrix} -R_{2,a} - k_a & 0 & k_b & 0 \\ 0 & -R_{2,a} - k_a & 0 & k_b \\ k_a & 0 & -R_{2,b} - k_b & 0 \\ 0 & k_a & 0 & -R_{2,b} - k_b \end{bmatrix} \quad \Lambda_L = \begin{bmatrix} -R_{1,a} - k_a & k_b \\ k_a & -R_{1,b} - k_b \end{bmatrix}$$

$$\Omega_T = \begin{bmatrix} -i\omega_z & 0 & 0 & 0 \\ 0 & i\omega_z & 0 & 0 \\ 0 & 0 & -i\omega_z & 0 \\ 0 & 0 & 0 & i\omega_z \end{bmatrix} \quad C_L = [R_{1,a} M_0^a \ R_{1,b} M_0^b]^T \qquad [9]$$

**EPG solution for evolution in absence of RF**

To move to the EPG representation we take the Fourier transforms of Eqs.[8] and write in terms of $\boldsymbol{F_n} = [\tilde{F}_n^a \ \tilde{F}_{-n}^{*a} \ \tilde{F}_n^b \ \tilde{F}_{-n}^{*b}]^T$ and $\boldsymbol{Z_n} = [\tilde{Z}_n^a \ \tilde{Z}_n^b]^T$:

$$\dot{F}_n = (\Lambda_T + \Omega_T) F_n \qquad [10]$$

$$\dot{Z}_n = \Lambda_L Z_n + C_L \delta(n) \qquad [11]$$

where it is understood that the full expression for the intra-voxel magnetization distribution consists of sums over *n* as in Eq.[1].

The solution to Eq.[10] is $\boldsymbol{F}_n(t + \Delta t) = \exp\big((\boldsymbol{\Lambda}_T + \boldsymbol{\Omega}_T)\Delta t\big)\,\boldsymbol{F}_n(t)$. We may take advantage of the fact that matrices $\boldsymbol{\Lambda}_T$ and $\boldsymbol{\Omega}_T$ commute in order to re-express the matrix exponential as a product of terms $\exp(\boldsymbol{\Lambda}_T \Delta t)\exp(\boldsymbol{\Omega}_T \Delta t)$ and define operators:

$$\boldsymbol{\Psi} \equiv \exp(\boldsymbol{\Omega}_T \Delta t) \quad [12]$$

$$\boldsymbol{\Xi}_T \equiv \exp(\boldsymbol{\Lambda}_T \Delta t) \quad [13]$$

such that $\boldsymbol{F}_n(t + \Delta t) = \boldsymbol{\Psi}\boldsymbol{\Xi}_T \boldsymbol{F}_n(t)$. Defining the dephasing during time $\Delta t$ as $\psi = -\omega_z \Delta t$ gives:

$$\boldsymbol{\Psi} = \begin{bmatrix} e^{i\psi} & 0 & 0 & 0 \\ 0 & e^{-i\psi} & 0 & 0 \\ 0 & 0 & e^{i\psi} & 0 \\ 0 & 0 & 0 & e^{-i\psi} \end{bmatrix} \quad [14]$$

which is the familiar 'shift' operator. This applies separately to each transverse state and simply increments the index as with the standard EPG algorithm. The operator $\boldsymbol{\Xi}_T$ accounts for both $T_2$ relaxation and exchange processes for transverse components.

The differential equation for longitudinal magnetization (Eq.[11]) is homogeneous for n≠0 but inhomogeneous for n=0. The solutions for the two regimes are:

$$\boldsymbol{Z}_n(t + \Delta t) = \boldsymbol{\Xi}_L\, \boldsymbol{Z}_n(t) \quad \text{(n≠0)} \quad [15]$$

$$\boldsymbol{Z}_0(t + \Delta t) = \boldsymbol{\Xi}_L\, \boldsymbol{Z}_0(t) + (\boldsymbol{\Xi}_L - \mathbb{I})\boldsymbol{\Lambda}_L^{-1}\boldsymbol{C}_L \quad \text{(n=0)} \quad [16]$$

$$\boldsymbol{\Xi}_L \equiv \exp(\boldsymbol{\Lambda}_L \Delta t)$$

This is in accordance with the standard EPG framework where longitudinal recovery occurs only in the *n=0* state. Note that the form of the $\boldsymbol{\Lambda}$ matrices means that exchange couples only states of the same *type* and *order* – for example $\tilde{F}_n^a \leftrightarrow \tilde{F}_n^b$ and $\tilde{Z}_n^a \leftrightarrow \tilde{Z}_n^b$.

**Solution for RF pulses**

Neglecting relaxation and exchange during RF pulses, their effect is to rotate the magnetization vector of each compartment independently (see refs. (24,25) for example). The general 3x3 rotation matrix acting on each compartment represented in the $[M_+\ M_-\ M_z]$ basis is (3):

$$T_{\alpha\phi} = \begin{bmatrix} \cos^2\frac{\alpha}{2} & e^{2i\phi}\sin^2\frac{\alpha}{2} & -ie^{i\phi}\sin\alpha \\ e^{-2i\phi}\sin^2\frac{\alpha}{2} & \cos^2\frac{\alpha}{2} & ie^{-i\phi}\sin\alpha \\ -\frac{i}{2}e^{-i\phi}\sin\alpha & \frac{i}{2}e^{i\phi}\sin\alpha & \cos\alpha \end{bmatrix} \quad [17]$$

where $\alpha$ and $\phi$ are the RF pulse flip angle and phase. Hence the overall transition matrix to apply to the full system $[\tilde{F}_n^a\ \tilde{F}_{-n}^{*a}\ \tilde{Z}_n^a\ \tilde{F}_n^b\ \tilde{F}_{-n}^{*b}\ \tilde{Z}_n^b]^T$ is simply:

$$T = \begin{bmatrix} T_{\alpha\phi} & 0 \\ 0 & T_{\alpha\phi} \end{bmatrix}. \quad [18]$$

The RF pulses do not mix the compartments, and no relaxation/exchange occurs in this time.

**Magnetization Transfer effects**

A different formulation is generally used when describing MT effects in tissues with a 'semisolid' component (18). In this case compartment *b* is often referred to as the 'bound' or 'restricted' pool and is assumed to represent highly immobile protons whose T$_2$ is very short (in the order of 10µs). In this case we assume that compartment *b* has no transverse magnetization, resulting in a complete system that is represented by four components where only longitudinal elements are coupled. In this system $M_T = [M_+^a\ M_-^a]^T$ and:

$$\Lambda_T = \begin{bmatrix} -R_{2,a} & 0 \\ 0 & -R_{2,a} \end{bmatrix} \quad \Omega_T = \begin{bmatrix} -i\omega_z & 0 \\ 0 & i\omega_z \end{bmatrix}. \quad [19]$$

In the proposed EPG framework this means that states $\tilde{F}_n^b$ and $\tilde{F}_{-n}^{*b}$ are dropped - we have a system with four states $[\tilde{F}_n^a\ \tilde{F}_{-n}^{*a}\ \tilde{Z}_n^a\ \tilde{Z}_n^b]^T$ per *n* value. The transverse magnetization is treated exactly as in the classic EPG case, in that it is subject to T$_2$ relaxation and shifts due to gradients. The coupled longitudinal magnetization states however evolve as per Eqs.[15] and [16].

The effect of RF pulses on compartment *a* is to rotate the magnetization as previously described. However for compartment *b* (the 'bound pool') RF pulses act so as to directly saturate the longitudinal component with saturation rate $\overline{W(\omega_z)}$ which for pulsed saturation is defined as (19):

$$\overline{W(\omega_z)} = \frac{\pi\gamma^2}{\tau_{rf}}\int_0^{\tau_{rf}} B_1^2(t)dt\ G(\omega_z). \quad [20]$$

$B_1(t)$ is the RF pulse waveform and $\tau_{rf}$ is its duration. This is a function of off-resonance frequency $\omega_z$ because it depends on the absorption lineshape $G(\omega_z)$. Different candidate lineshapes have been proposed for modelling semisolids in biological tissues with Gaussian (19) and Super-Lorentzian lineshapes (26) used primarily. The overall RF transition matrix is therefore

$$T = \begin{bmatrix} T_{\alpha\phi} & 0 \\ 0 & e^{-\overline{W(\omega_z)}\tau_{rf}} \end{bmatrix} \quad . \qquad [21]$$

**Summary of proposed theory**

To summarize, we have introduced extensions to the EPG formalism to account for multi-compartment systems with exchange, and refer to this as EPG-X. There are two variants – one for systems governed by the Bloch-McConnell equations denoted BM, and one for the variant of BM often used for MT, in which one compartment has negligible transverse magnetization. Both effectively use two coupled EPG calculations, one for each compartment, however the MT variant uses a reduced second compartment with longitudinal components only. These are summarized diagrammatically on Figure 1.

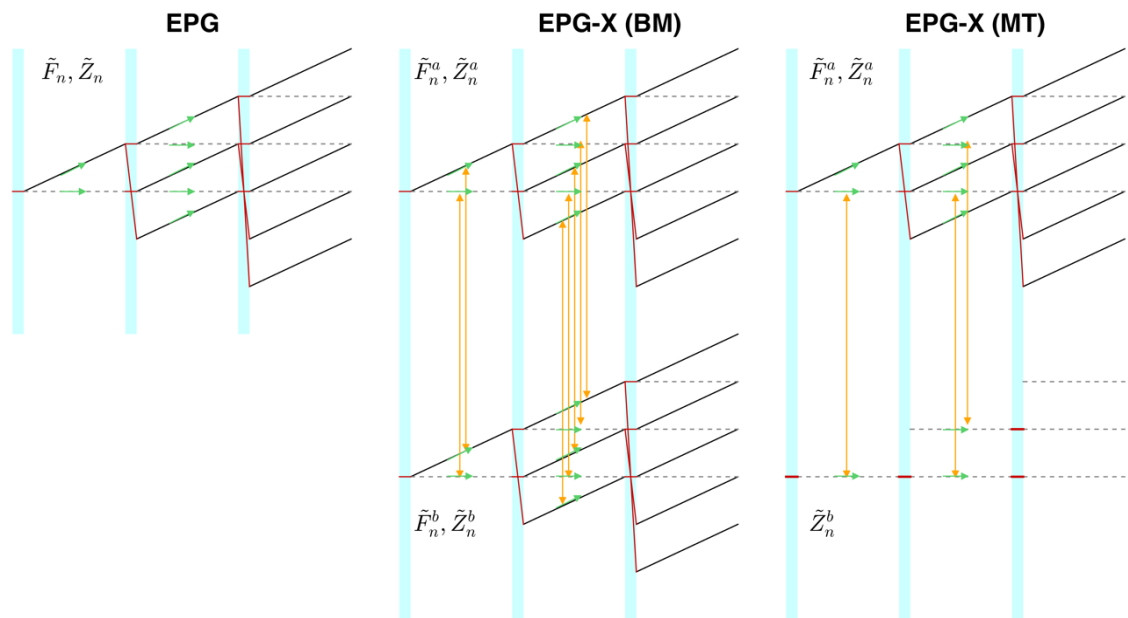

*Figure 1: **Left**: overview of 'classic' EPG algorithm. $\tilde{Z}_n$ states are represented by dotted lines, $\tilde{F}_n$ by solid black lines. Blue shaded regions correspond to application of RF pulses, red lines trace how states are mixed by the action of RF pulses. Green arrows depict relaxation effects, which occur individually to each state during the period between RF pulses. **Middle**: EPG-X (BM) approach consists of two separate EPGs. RF pulses have the same effect as EPG. During evolution periods relaxation (green arrows) and exchange (yellow arrows) both occur. Exchange links each state to its equivalent in the other compartment, i.e. $\tilde{F}_n^a \leftrightarrow \tilde{F}_n^b$ etc. **Right:** EPG-X (MT) has a reduced second compartment represented only by $\tilde{Z}_n^b$. RF pulses directly saturate these states (Eq.[21]) and they exchange directly with their equivalents in compartment a. For clarity, on all diagrams relaxation and*

exchange effects are only depicted for the periods following the first two RF pulses. Note that although depicted by separate arrows, relaxation and exchange processes are governed by single combined operators.

|  |  | $T_{1,a}$ /ms | $T_{1,b}$ /s | $T_{2,a}$ /s | $T_{2,b}$ /ms | $k_a$ / s$^{-1}$ | f |
|---|---|---|---|---|---|---|---|
| Generic model | BM | 1000 | 500 | 100 | 20 | 10 | 0.2 |
|  | MT | 1000 | 1000 | 100 | 12x10$^{-3}$ † | 10 | 0.2 |
| White matter | MT | 779 | 1000 | 45 | 12x10$^{-3}$ † | 4.3 | 0.117 |

Table 1: Model parameters used in simulation experiments. By convention compartment a is larger and f is the fractional size of compartment b. $k_a$ is the exchange rate from a to b. The 'white matter' model uses median parameters estimated for 1.5T by Gloor et al (table 1, ref (27)). †MT experiments used a Super Lorentzian absorption lineshape model with $T_{2,b}$=12µs as described in the text, giving G(0)=15.1 µs

**Methods**

EPG-X is a general framework that could in principle be used to simulate the response of a two-compartment system for any pulse sequence. The theory is illustrated by simulating four separate scenarios, linking to existing solutions where available. The tissue parameters used in the following simulations are outlined in Table 1; the subset used in each simulation is indicated in the text. The 'generic model' uses round numbers to aid interpretation of results. The 'white matter' model is an MT based model taken from median measured white matter properties at 1.5T from Gloor et al (table 1, ref (27)). The Super-Lorentzian lineshape function has been used for $G(\omega_z)$; as in ref (27) the function was extrapolated between ±1kHz by fitting a spline to avoid the singularity at zero frequency. In both models, compartment *a* is the larger one; for MT compartment *b* represents bound (macromolecular) protons.

All numerical simulations and analysis were performed using Matlab R2015a (The Mathworks, Natick, MA). A fully functional implementation is available to download at http://www.github.com/mriphysics/EPG-X . Code for generation of all simulations presented in this paper is included - hash 7da7f9e was the version at time of submission.

**Test 1: Steady State solutions for Gradient Echo sequences**

Steady-state expressions for spoiled gradient echo (SPGR) and balanced steady-state free precession (bSSFP) sequences have been derived for BM and MT models in terms of the matrices already introduced in the theory section – these are given in the Appendix. These solutions have been provided elsewhere in the literature. The BM versions were used in the multi-component DESPOT method - see equations 1 and 4 in ref. (22). For the MT case the SPGR solution appears for example in ref (28); extension to bSSFP is straightforward (29). An analytic solution for the bSSFP case has also been presented for on-resonance magnetization (27). The SPGR solutions assume perfect spoiling of transverse magnetization and reduce to the Ernst formula in the single compartment case.

The steady-state solutions outlined were tested against the EPG-X predictions for the 'generic' model parameters listed in Table 1. For comparison single component (classic EPG) calculations were performed using T1=1000ms, T2=100ms. Sequences with TR=10ms and flip angle ($\alpha$) 15° were simulated for 500 TR periods (i.e. 5xT$_1$) after which a steady state was assumed to have formed. For SPGR, EPG simulations were repeated for different values of the quadratic RF spoiling phase increment $\Phi_0$ from 0° to 180° in steps of 0.5°. RF saturation term $\overline{W}$ was computed by assuming hard pulses with maximum amplitude 13.5$\mu T$.

Equation 1 implies that $M_+(\psi)$ and $M_z(\psi)$ may be obtained from the EPG predictions by performing an inverse FFT over 'order' parameter *n* (30). For bSSFP case since net gradient area is zero, $\psi$ is simply the phase gained due to off-resonance effects in a given TR period. Hence we may derive the familiar off-resonance sensitivity of the bSSFP method by applying iFFT to the EPG predictions. These were compared with the steady-state solutions Eq.[A2].

**Test 2: MT in Transient Gradient Echo Sequences**

Magnetization Transfer effects have been shown to strongly affect the SPGR signal in the steady state (28,31). Transient gradient echo sequences with variable flip angle, often following inversion pulses, have been used for MRF (21,32), hence we simulated a simple example of such a sequence to predict transient behaviour. The sequence employed an adiabatic inversion pulse followed by a series of 256 low flip angle RF pulses whose amplitude was varied sinusoidally (shown in results section); some pulse amplitudes were zero to allow for magnetization recovery. The RF pulse energies were 433 $ms\ \mu T^2$ for the inversion and 54.3 $\alpha^2\ ms\ \mu T^2$ for the small flip angle pulses ($\alpha$ is the flip angle, rad). A constant repetition time of 12ms was used with constant gradient area in each TR period, even those with zero flip angles. The sequence was simulated with EPG-X (MT) using white matter parameters from Table 1. Both bSSFP and SPGR were simulated using the same timing.

Systems with MT effects have been shown to exhibit bi-exponential T1 recovery (31,33) – for the two pool model the $T_1$ observed from inversion recovery measurements may be related to the MT parameters as follows (31):

$$T_1^{obs} = \left\{ \frac{T_{1,a}^{-1}+k_a+T_{1,b}^{-1}+k_b}{2} - \frac{\sqrt{(T_{1,a}^{-1}+k_a+T_{1,b}^{-1}+k_b)^2 - 4(T_{1,a}^{-1}T_{1,b}^{-1}+T_{1,a}^{-1}k_b+T_{1,b}^{-1}k_a)}}{2} \right\}^{-1} \quad [22]$$

For the white matter model Eq.[22] yields $T_1^{obs}$=799ms. For consistent comparison, single compartment EPG was simulated using $T_1^{obs}$ and $T_{2,a}$ since these are the parameters that would be measured using standard inversion recovery and spin echo methods.

**Experimental measurements**

Physical experiments were performed on a Philips (Best, Netherlands) Achieva 3T MRI scanner. Two phantoms were made from 25ml sample tubes: water doped with 0.1 mM $MnCl_2$ (expected to have no MT effect), and cooked egg white which is expected to exhibit a measureable MT effect (34). The samples were imaged using the SPGR sequence described above (TR=12ms, TE=2.9 ms) with frequency encoding aligned with the longitudinal axis of the tubes and phase-encoding switched off in order to directly record echo amplitudes. Experiments were repeated with RF spoiling phase increment $\Phi_0$ set to 150° (default) and 117°. Single compartment relaxation times and apparent diffusion coefficient (*D*) were measured for each phantom, using inversion recovery TSE ($T_1^{obs}$), multi-echo spin echo ($T_2$) and diffusion weighted spin echo (*D*) respectively. For the water phantom $T_1^{obs}$=899±5ms, $T_2$=92±2ms, and D=2.35±0.18x$10^{-3}$mm$^2$s$^{-1}$; for egg white $T_1^{obs}$=1577±23ms, $T_2$=96±5ms, and D=1.92±0.05x$10^{-3}$mm$^2$s$^{-1}$. Separate long TR multi-flip angle measurements were made in order to precisely measure the effective $B_1$ amplitude and $M_0$ (including receiver coil scaling) in these phantoms; these measurements were used to match the measured echo amplitudes as closely as possible to EPG simulations.

As will be shown later, it was found that diffusion effects needed to be taken into account to accurately model the SPGR sequence. Diffusion can readily be accounted for in the EPG framework (12) - essentially it attenuates higher order states by destroying coherence at small length scales. We followed the methodology used by Weigel to implement this in the single compartment EPG model. Further, we experimented with extending this to multi-compartment models by applying the same treatment independently to each compartment – validity of this approach is discussed later. Note that for results presented in this paper, diffusion effects were only included for those relating to this experiment.

**Test 3: Chemical Exchange in multicomponent T2 analysis of CPMG data**

Multicomponent analysis of multiecho CPMG spin echo data is a well-known method for estimation of myelin fraction in white matter (35). If perfect 180° refocusing pulses are assumed, the multiecho signal can be analysed using non negative least squares (NNLS) fitting to an exponential model (36). However, $B_1^+$ inhomogeneity effects will introduce other stimulated echoes that make the data deviate from this simple model. Prasloski et al (16) showed that an EPG-based model may be used instead and postulated a multi-component model with no exchange effects between components. In order to explore any potential effects from exchange we used the EPG-X (BM) model for a two-component system to simulate multiecho data for a range of exchange rates and $B_1^+$ scaling factors. In each case the simulated data was analysed using the classic NNLS fitting approach from which the estimated small pool fraction was taken as the area of smaller peak in the T2 'spectrum'.

Simulations all used 50 echoes with spacing 5ms. The generic model parameters from Table 1 were used, except for $k_a$ which was varied from 0 to 2.5s$^{-1}$ – for $f$=20% this corresponds to the reverse exchange rate $k_b$ varying from 0 to 10s$^{-1}$. The 1/$k_b$ is the mean residence time in the small pool (i.e. myelin water residence time if compartment $b$ is myelin water). $B_1^+$ scaling factors from 0.75 to 1.25 were included. NNLS was performed using Matlab function *lsqnonneg*.

**Test 4: Magnetization Transfer effects in multislice TSE imaging**

Multislice TSE is also sensitive to MT effects (37,38) since from the point of view of a given slice location, the acquisition of the other slices may be viewed as repeated off-resonant irradiation. Melki and Mulkern (37) proposed a semi-empirical model to quantify these effects and Weigel et al (38) extended this to consider TSE using low refocusing angles. We used the EPG-X (MT) framework to model a multislice TSE sequence with five slices, similar to that in ref (38), in white matter with properties as listed in Table 1. Off-resonant excitation of other slices can be modelled by trains of pulses with zero flip angle for compartment $a$ but with saturation still applying to compartment $b$. The sequence had 27 echoes, interecho spacing 7.3ms and TR=5000ms. Slices were spaced by 2kHz and were assumed to not affect each other's free magnetization (i.e. compartment $a$). The five slices had frequency offsets -4,-2,0,2,4 kHz; absorption values were $G(0) = 15.1\mu s$, $G(\pm 2kHz) = 10.8\mu s$ and $G(\pm 4kHz) = 6.7\mu s$. To simulate different RF pulse shapes, three different values for the pulse energy $<B_1^2>$ were used: 20 $\alpha_{RMS}^2$, 30 $\alpha_{RMS}^2$ and 40 $\alpha_{RMS}^2$ $ms\,\mu T^2$ ($\alpha_{RMS}$ is the RMS flip angle in radians). The simulations were done from the point of view of the slice at the centre of the group (i.e. slice 3) with acquisition interleaved in order 1,3,5,2,4. Simulations were run for three TR periods to ensure equilibrium was reached, and a range of refocusing flip angles from 50° to 180° were used. Single slice imaging with the same TR was simulated for comparison.

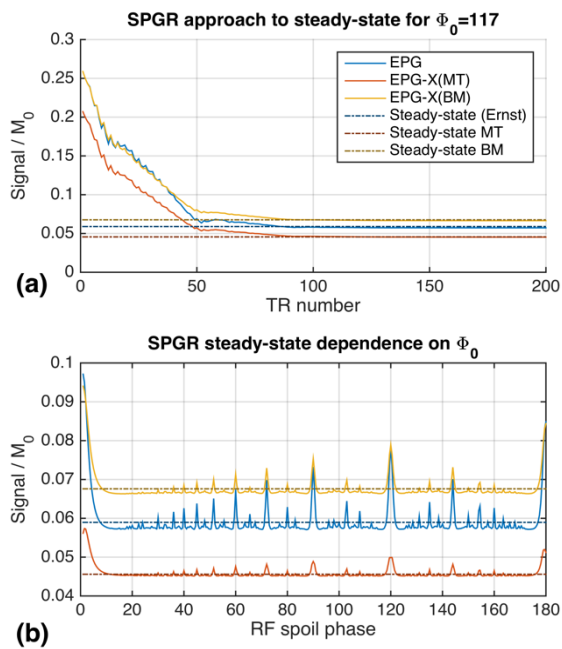

*Figure 2:* (a) Approach to steady-state for SPGR sequence with TR=10ms, $\alpha = 15°$ computed with EPG, EPG-X (BM) and EPG-X (MT) for 'generic model' parameters (see Table 1). Expected steady-state signals were computed using Eq.[A1]. After around 100 TR periods the different transient simulations each approach the expected steady-state. (b) Steady state signal as a function of RF spoiling phase increment $\Phi_0$. The steady state signals for the single compartment, MT and BM models are all different. All are variable with $\Phi_0$ and as expected do not always agree with the direct steady state calculations which are computed assuming perfect spoiling.

## Results

**Test 1: Comparison with existing steady-state gradient echo solutions**

Figure 2a shows the approach to steady-state for SPGR with $\Phi_0 = 117°$ for standard EPG and the proposed variants, compared with the ideal spoiling steady-state values predicted by Eq.[A1]. All three curves approach the ideal spoiling steady-state after about 100 TR periods. Figure 2b shows the steady-state values reached by the EPG methods for a range of $\Phi_0$ compared with the ideal spoiling prediction – as expected strong variation with $\Phi_0$ is seen. The degree of variation is also different for each model – perhaps surprisingly the MT case shows much less variability than the others. The EPG-X predictions broadly agree with the directly calculated steady states but give additional information by properly characterising the effect of RF phase cycling without assuming ideal spoiling.

Figure 3 compares EPG and steady-state solutions for bSSFP. The left column shows the familiar off-resonance sensitivity profile as it evolves through time; this is obtained by performing iFFT on the

EPG data as described by Eq.[1]. The right column compares the steady-state profiles predicted from theory (Eq.[A2]) with the EPG predictions after 500 TR periods when a steady-state has definitely been reached. Each model produces quite different steady-state behaviour, however in each case the agreement between EPG-X and direct calculation is excellent, as is agreement with the on-resonance analytic expression for the MT case (red asterisk, ref (27)).

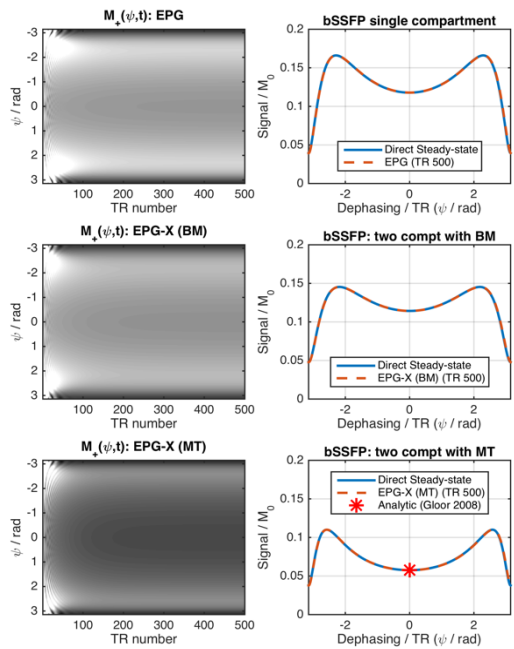

*Figure 3: Left column:* Approach to steady-state as a function of off-resonance parameter $\psi$ for bSSFP computed for standard EPG, EPG-X (BM) and EPG-X (MT); off-resonance behaviour is obtained by inverse FFT, see text for details. *Right column:* Steady-state solutions from Eq.[A2] compared with EPG based predictions after 500 TR periods. There is a very close correspondence between the expected steady-state and the value reached by each of the EPG calculations. For EPG-X (MT) there is also agreement with the analytic solution derived for $\psi=0$ in ref (27).

**Test 2: 'MRF' style transient gradient echo**

The variable flip angle profile used is illustrated in Figure 4a. The figure also illustrates the expected signals (Fig.4b) and evolution of longitudinal magnetization (unmodulated $\tilde{Z}_0$ states, Fig.4c) in the white matter model for the SPGR sequence, comparing EPG-X with single pool EPG using $T_1^{obs}$ (i.e. the $T_1$ that would be measured for this system using inversion recovery). The signal profiles are different, particularly straight after the inversion (at the beginning of the sequence) when the magnetization in the MT system recovers more quickly.

Figure 5 shows the equivalent result for a balanced SSFP sequence – as with test 1 the full off-resonance sensitivity profile was recovered by inverse FFT, and line profiles are plotted for $\psi = 0$ and $\psi = \pi/2$. Similar behaviour is observed to the SPGR case, with larger discrepancy in the early part following the inversion. Note that the influence of MT appears to be off-resonance dependant (a function of $\psi$) – this is not related to the RF pulse bandwidths since instantaneous pulses imply infinite bandwidth.

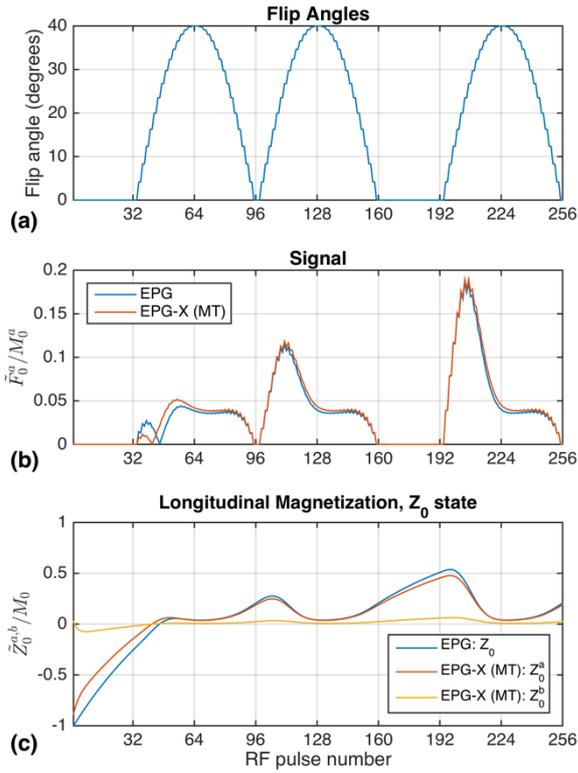

*Figure 4: (a) Variable flip angle train used for numerical and physical experiments. There are 16 different flip angles including zero which is included to allow some free recovery of magnetization. These pulses immediately follow an adiabatic inversion pulse (index 0). (b) Predicted signals from EPG and EPG-X (MT) model for white matter parameters (Table 1). Note that the EPG-X signals are normalised by (1-f) to account for compartment b being 'invisible'. MT leads to different behaviour, particularly immediately after the inversion pulse. (c) $\tilde{Z}_0^{\square}$ from the EPG and EPG-X models. For EPG-X the level of saturation of compartment b changes dynamically, leading to altered dynamics of the observed signal when compared with a single compartment model.*

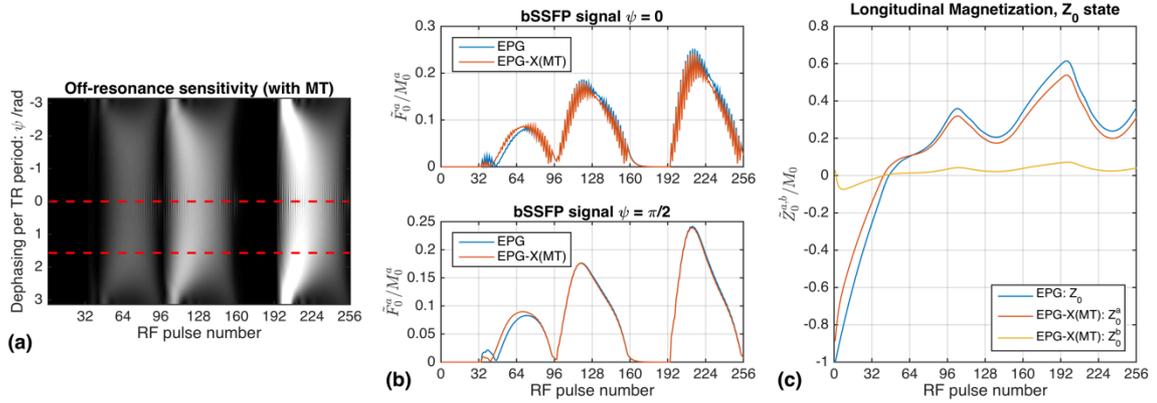

*Figure 5: (a) Signal as a function of TR number and $\psi$ for bSSFP. (b) Profiles for $\psi = 0$ and $\psi = \pi/2$ (see dotted lines on (a)). Note that the oscillations in the $\psi = 0$ case are due to the 'stepped' nature of the changing flip angles (see Figure 4a). The effect of MT alters the signal dynamically, particularly after the inversion as with SPGR (Fig.4). The difference between EPG and EPG-X also changes as a function of off-resonance parameter $\psi$. (c) $\tilde{Z}_0$ profiles; for EPG-X the saturation of compartment b varies dynamically.*

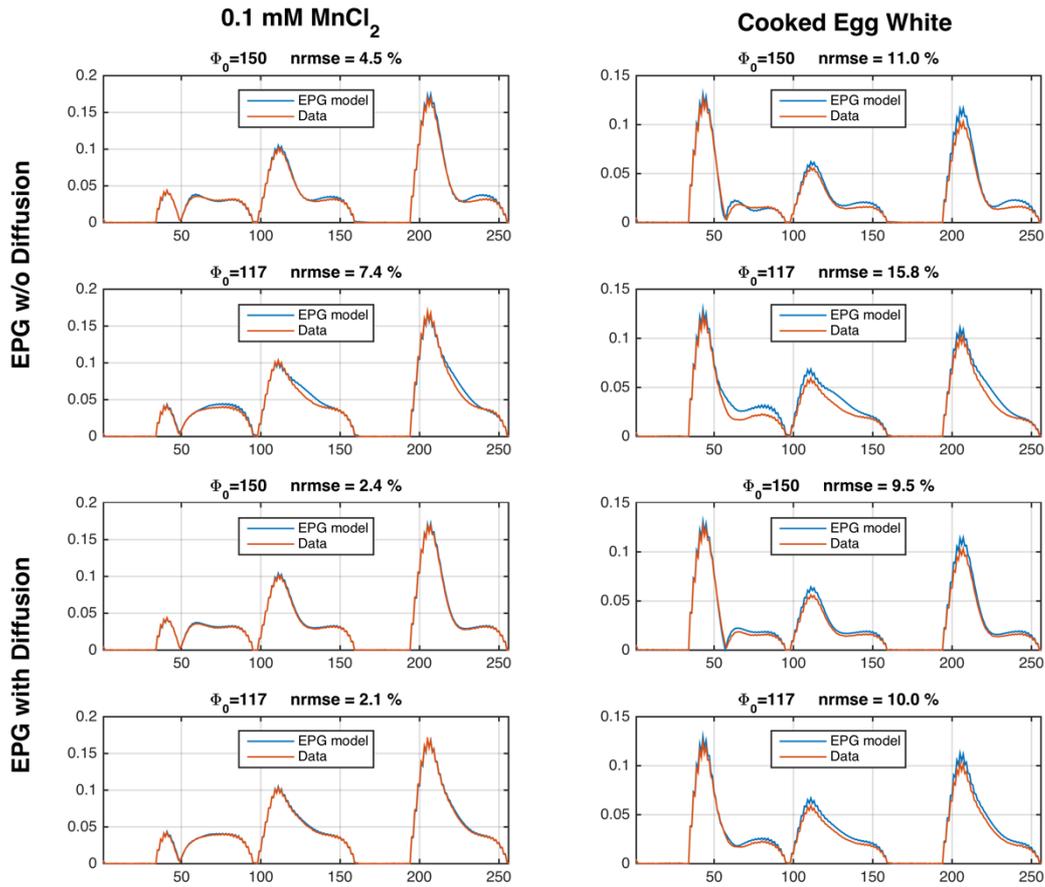

*Figure 6  Experimental SPGR data compared with classic single compartment EPG model. No fitting was performed – relaxation times, B1 scaling and receiver-weighted $M_0$ scaling factors were experimentally measured. When diffusion is not included (**top two rows**) the match to EPG is not perfect. For the $MnCl_2$ phantom once diffusion is included (**bottom two rows**) the match is very good (nrmse~2%). For the egg white phantom there remain systematic differences, suggesting that the single compartment model is not sufficient. Note also that the observed signal profiles are quite different for the two different values of $\Phi_0$ as predicted by the EPG model.*

Figure 6 compares experimentally obtained SPGR data with standard EPG predictions; the data and model are not fitted together; all necessary parameters and scaling coefficients were measured in calibration experiments. For the $MnCl_2$ phantom, reasonable agreement is obtained using normal EPG however this is significantly improved by including diffusion effects as described in ref (12). Note also how the signal profiles from the two different RF spoiling phase increments $\Phi_0$ are quite different. For egg white the match to classic EPG is also improved by adding diffusion effects, however there are systematic differences which we hypothesized might be due to MT effects. The experimentally obtained data for egg white were fitted to the EPG-X (MT) model by minimising the mean square deviation by optimizing over $T_{1,a}$, $T_{1,b}$, $k_a$, and G(0). Fitting used Matlab function *fmincon* including a non-linear constraint enforcing consistency between estimated parameters and measured $T_1^{obs}$ (via Eq.[22]). Yeung and Swanson also studied heat denatured hen egg albumen with a two compartment MT model (39); following their approach we fixed *f*=0.082 since this was a reported literature value from wet/dry weight measurements on hen eggs and constrained $T_{1,a}$ to be close to 3s. Data for $\Phi_0 = 150°$ and $\Phi_0 = 117°$ were fit simultaneously. $T_{2,a}$, D, and overall scaling constant were fixed at the measured values. Diffusion was implemented in EPG-X using the same approach as for standard EPG, the validity of this approach will be discussed later. Figure 7 shows the fit that could be obtained – there is good but not perfect agreement using the following parameters: $k_a$= 1.11s$^{-1}$, $T_{1,b}$=222ms $T_{1,a}$=2801ms, G(0)=43μs. This combination of parameters would yield $T_1^{obs}$=1605ms, which is approximately one standard deviation away from that measured by inversion recovery.

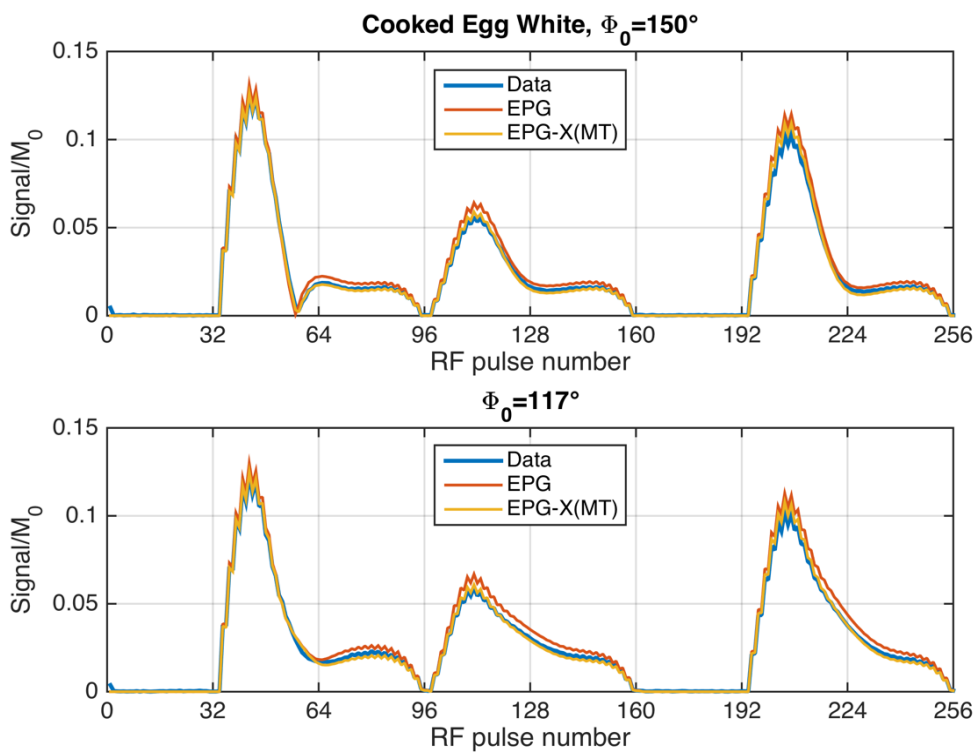

**Figure 7** *Result of fitting the egg white phantom data to the EPG-X (MT) model; data from both values of $\Phi_0$ were fit simultaneously. $T_2$, diffusion coefficient and other scaling factors were held fixed at experimentally measured values, and f=0.082 was taken from literature. Following ref (39) $T_{1,a}$ was constrained to be close to 3s. $k_a$, $T_{1,a}$, $T_{1,b}$ and G(0) were varied. Best fit parameters were $k_a$= 1.11s$^{-1}$, $T_{1,a}$=2800ms, $T_{1,b}$=222ms and G(0)=43μs. These would yield $T_1^{obs}$=1605ms. NRMSE before fit = 9.5%, after fit = 3.9%.*

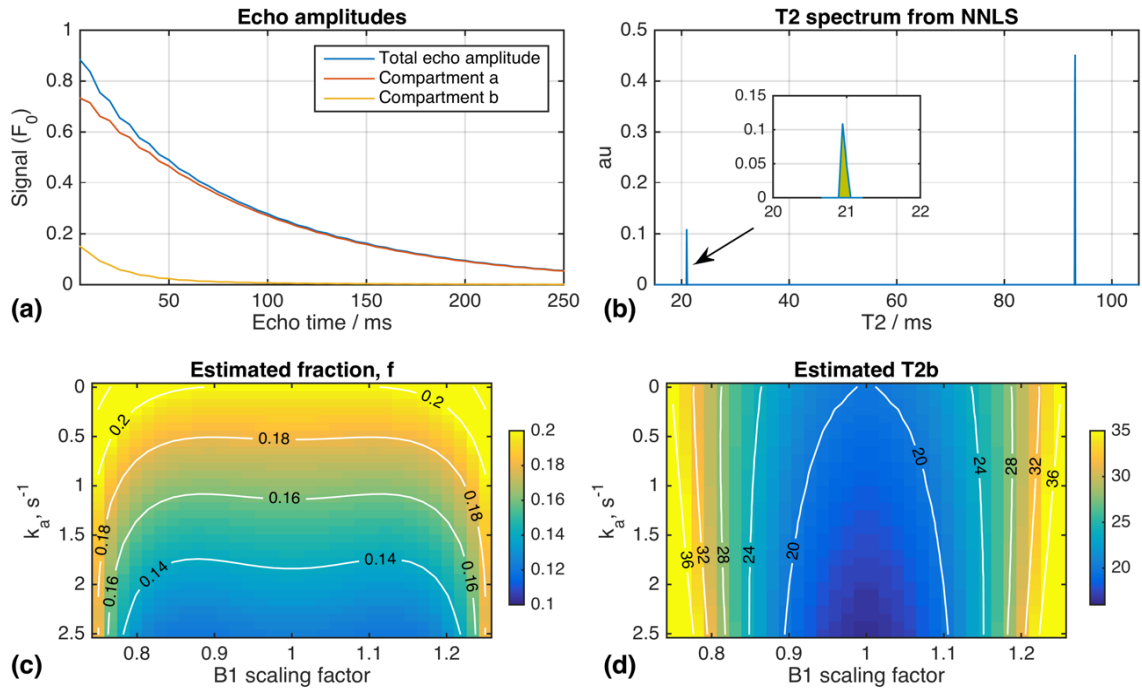

*Figure 8* Results from test 3. *(a)* Example echo amplitudes for $k_a=1s^{-1}$ $B_1$ scaling = 1.05; both compartments are shown but only the total echo amplitude would be observed experimentally. *(b)* T2 spectrum from data in (a) obtained using NNLS. Inset plot shows shorter T2 peak; estimated $T_{2,b}$=20.9ms, f=0.163 (obtained from peak area, shaded). *(c)* Estimated fraction as a function of $k_a$ and $B_1$ scaling – f is mainly a function of $k_a$. When $k_a$=0 f is estimated correctly, otherwise f tends to be systematically underestimated. *(d)* Estimated $T_{2,b}$ – while there is some variation with $k_a$ this is mainly a function of $B_1$ scaling.

**Test 3: Multiecho CPMG relaxometry**

Figure 8a shows example echo amplitudes from the simulated multiecho CPMG data ($B_1$ scaling = 1.05, $k_a$=1s$^{-1}$); Fig.8b has the corresponding $T_2$ spectrum from NNLS analysis. There are two peaks corresponding to the two compartments ($T_{2,a}$=100ms,$T_{2,b}$=20ms) and the fraction *f* is estimated by taking the ratio of the peak areas (shaded on Fig.8b). Fig.8c and 8d show variation in estimated *f* and $T_{2,b}$ with $k_a$ and $B_1$ scaling; the apparent $T_{2,b}$ varies strongly with the $B_1$, whereas the apparent *f* is more strongly dependent on exchange rate $k_a$. If exchange is present then *f* tends to be underestimated. If this simple model system were to represent myelin water exchanging with intra/extra-axonal water then a reasonable value for myelin-water residence time might be 200ms (although estimates vary significantly with anatomy (40)). This would give $k_b$=5s$^{-1}$ and hence $k_a$=1.25s$^{-1}$ for *f*=0.2, for which Figure 8c indicates the estimated fraction would be 0.156; a 22% underestimate.

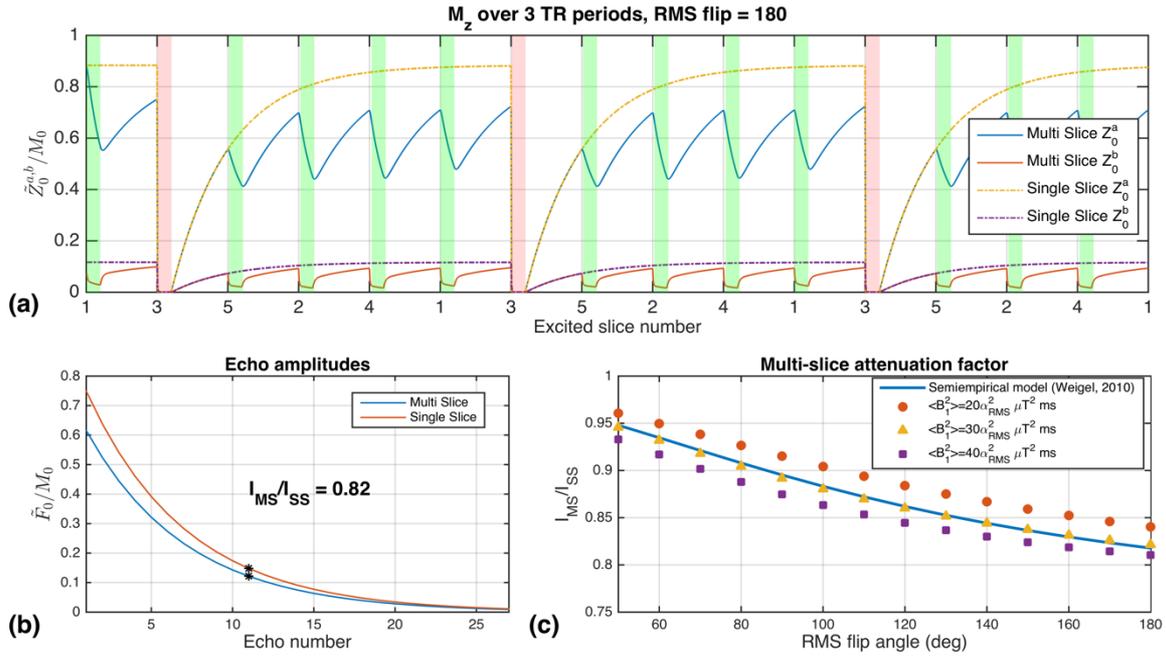

*Figure 9: Comparison of multislice with single slice TSE for MT white matter model. Simulated sequence had 5 slices, these results focus on slice 3 (central slice). (a) Evolution of $\tilde{Z}_0^a$ ('free pool') and $\tilde{Z}_0^b$ ('bound pool') for slice 3 during multislice and single-slice TSE. Red shaded areas show when slice 3 is being acquired, green areas are when other slices are being acquired in multislice case. For the multislice $\tilde{Z}_0^b$ is saturated by off-resonant excitation and this causes a saturation of $\tilde{Z}_0^a$ via magnetization transfer. Dotted lines illustrate recovery in single-slice case when only slice 3 is acquired. (b) Echo amplitudes acquired in third TR period – the multislice exam has lower signals because of the saturation from off-resonant slice excitation. (c) Attenuation factor of multislice compared to single slice over range of RMS flip angles and pulse energies, compared with semiempirical model from ref (38)*

**Test 4: Multislice TSE imaging**

Comparison of single slice with multislice TSE is shown in Figure 9. The shaded regions on part (a) correspond to times when RF is being transmitted; red shading indicates the slice of interest being excited (slice 3) and green shading indicates other slice acquisitions for the multislice case. The plotted longitudinal magnetization corresponds to slice 3 – off-resonant slice excitation leads to strong saturation of on-resonant magnetization as expected. Saturation is less severe when slices 1 and 5 are excited (these are the furthest away, offset ±4kHz). Fig.9b illustrates the echo amplitudes obtained - significant loss (~18%) is seen for the 180° refocusing pulse. EPG-X may be used to simulate the full dynamics of the signal for any flip angle. A more limited semiempirical model that considers only the average RF power can be used to characterise the signal loss, and Weigel et al fitted this model using two-parameters to measurements on white matter (see Figure 2 ref (38)).

Fig.9c compares the semiempirical model with the EPG-X predictions, and shows both predict a similar trend in increasing attenuation of the multislice signal with flip angle. The RF pulse energy used in ref (38) was not specified; we tried three plausible values and found that 30 $\alpha_{RMS}^2$ would give excellent agreement between the EPG-X simulations and the semiempirical model.

**Discussion**

This work has introduced a general framework modelling of processes governed by the Bloch-McConnell (BM) equations, or their modified form for magnetisation transfer (MT) using Extended Phase Graphs. Essentially the different compartments are described by separate phase graphs, which exchange with each other during evolution periods. For the MT case the 'bound' protons have no transverse components so their phase graph consists only longitudinal states. No fundamentally new biophysical model has been proposed in this work, rather we have shown how existing methods can be incorporated into the EPG framework, which may then be used for efficient simulation and/or intuitive analysis of sequence behaviour. The new model, referred to as EPG-X, has been validated by comparing steady-state behaviour for gradient echo sequences with existing solutions. The EPG-X (MT) model was also found to give results consistent with prior work on MT effects in multislice TSE (38). In both of these examples, while the EPG-X calculation agrees with the literature it provides a richer solution; yielding the approach to steady-state and effect of RF spoiling for the gradient echo simulations, and allowing prediction of expected signal attenuation without requiring empirical measurement of sequence dependent parameters for the TSE.

There has been particular recent interest in using EPG for simulations as part of relaxometry and other quantitative MR approaches; in this work we focused on two examples - multiecho CPMG and non-equilibrium gradient echo sequences of the type used in MRF. Simulating the CPMG data using the EPG-X (BM) framework and then analysing using standard NNLS methods, we found that while $B_1$ errors lead primarily to overestimation of $T_{2,b}$, exchange processes lead to an underestimate of the small compartment fraction. Exchange processes have been suggested in the literature as a potential reason for variable measurements of myelin-water fraction using CPMG methods on rat spinal cord (41).

For the MRF style sequence we focused on the possible impact of MT effects in white matter – this has been shown to interfere with traditional gradient echo based relaxometry approaches, for example in ref. (28). Figures 4 and 5 show that MT will lead to altered behaviour that is particularly pronounced after inversion, for both bSSFP and SPGR sequences. The reason can be seen in the plots of $\tilde{Z}_0$ in Figs.4 and 5 - in the EPG-X (MT) model the 'free' magnetization in compartment *a* recovers quickly by receiving a transfer of magnetization from 'bound' compartment *b*. Once *b* is inverted, *a* starts to recover more slowly. Measurements on doped water and egg white phantoms (Fig.6) also suggest that while the former can be well modelled using standard EPG (excellent agreement is obtained with no fitting to the data), the residuals are larger for the latter. Even for the water phantom good agreement was only obtained once diffusion is included in the EPG model. Also note the measured signal profiles from using different RF spoiling phase increments are quite different. Variability with RF spoiling can also be seen in the steady-state simulations (Figure 2b) and is an important consideration for steady-state relaxometry methods (42). Interestingly Figure 2b suggests that MT effects may dampen the apparent variability of the SPGR signal with phase increment $\Phi_0$.

The types of deviation between EPG and EPG-X models shown in Figs.4 and 5 might be a source of error for MRF, however since the effects are subtle and variable through time they may also prove to be incoherent with the generated dictionaries – this remains to be investigated. The fact that the observed 'fingerprint' profiles are affected by MT could also imply that there is potential for quantitative MT characterisation with this type of sequence. The saturation of compartment *b* ('bound protons') varies dynamically throughout these sequences because the RF power is constantly changing – potentially a means for probing these effects. A preliminary study on use of MRF with a BM model has shown some promise (43). In order to demonstrate this potential we used non-linear fitting to estimate the MT specific parameters (Figure 7). The fitted parameters ($k_a$=1.11s$^{-1}$, $T_{1,a}$=2801ms, $T_{1,b}$=222ms) were similar to those measured in egg white by Yeung and Swanson (39) ($k_a$=3s$^{-1}$, $T_{1,a}$=3s, $T_{1,b}$=190ms; made at 2.0T) and were in agreement with $T_1^{obs}$ from inversion recovery. We followed the approach in ref (39) of constraining $T_{1,a}$ to be close to 3s, but better agreement could be obtained by relaxing this constraint, yielding very different values for the other parameters (data not shown). Clearly more work is needed to determine accuracy and validity of using transiently varying MRF style sequences for quantifying MT related parameters, however our results show that a simultaneous fit to two acquisitions with different RF spoiling parameters can be obtained with reasonable parameter estimates.

Multi-compartment models seek to explain complex underlying biological systems, but choice of model is key. White matter is a particularly complex tissue; in this paper 'white matter' was modelled using a two-compartment MT approach in tests 2 and 4, while test 3 which relates to $T_2$ relaxometry (also commonly used for studying brain tissue) used a BM model. This reflects the range of models that currently exist in the literature; choice of model depends on the type of sequence being modelled as well as the tissue. For multiecho $T_2$ relaxometry the BM model is most relevant since it is seeking to measure multiple compartments with appreciable $T_2$; for evaluating the effect of off-resonant saturation in multislice imaging the MT model is most relevant. More complex mixed models that include multiple BM and MT compartments (24,44) have also been proposed; Liu et al have proposed a general framework for such systems (25). In this work we focused on two compartment models, but in principle the same arguments used in ref (25) can be used to extend the EPG-X formalism to more compartments as well. The cost of doing so is increased complexity and more (perhaps difficult to estimate) parameters that must be specified.

**Implicit assumptions and validity**

EPG-X assumes that the underlying magnetization in both compartments forms a spatial distribution at a sub-voxel length-scale. Exchange couples Fourier configurations in one compartment with the same configuration in the other compartment – this is equivalent to assuming that exchange interactions couple these distributions *locally,* i.e. the magnetization in compartment *a* at one location couples to compartment *b* at the same location, but not adjacent locations. 'Isochromat' based modelling methods (e.g. refs (25,45)) make the same assumption in the spatial domain. This is physically reasonable since both chemical exchange and MT occur at the level of individual molecules, far smaller than any scale that is of interest for modelling the sub-voxel magnetization distribution at spatial resolutions relevant to MRI.

It was also found that diffusion effects must be accounted for to accurately match observed signals to EPG models. As far as we are aware there is no commonly adopted equivalent model for the multi-compartment case, since we are effectively combining the Bloch-McConnell and Bloch-Torrey equations. For the measured egg white data (Figure 7) we took the most basic approach, which was to treat diffusion effects independently for both compartments but with the same diffusion coefficient for each. It might be expected that actually diffusion coefficients would be quite different for each compartment and this could readily be achieved within the same framework. Further work is needed to identify the most appropriate biophysical model.

Finally, this work has been presented in terms of the 'regular time increment' version of the EPG framework which allows configuration states are considered in terms of integer indices only. Sequences with variable gradient directions and/or non-uniform timing are described instead using a continuous Fourier transform, see ref (3) for a detailed discussion. The theory put forward in this paper would generalise readily to this approach since none of the exchange-related operators are explicit functions of space.

**Conclusions**

Extensions to the EPG framework to systems governed by the Bloch-McConnell equations and modified forms for pulsed MT have been proposed. The new formalism named EPG-X may be used to efficiently model a wide range of pulse sequences and results indicate that for steady-state sequences EPG-X gives equivalent predictions to commonly used solutions. EPG-X could prove to be useful for quantitative imaging, particularly for non-steady-state sequences where accurate modelling of the transient response is necessary.

## Appendix: Steady-state solutions for gradient echo sequences

### Spoiled Gradient Echo

For an 'ideally spoiled' sequence we consider only a steady-state formed by longitudinal components. The measured signal is given by:

$$M_+^{ss} = \Theta\{\mathbb{I} - \Xi_L T_L\}^{-1}\{\Xi_L - \mathbb{I}\}\Lambda_L^{-1} C_L \qquad [A1]$$

where time increment $\Delta t = TR$ is used in the definition of $\Xi_L$. Operator $\Theta$ represents the excitation and sampling of longitudinal magnetization, and is defined differently for the BM and MT scenarios:

BM case: $\qquad \Theta = [\sin\alpha \; \sin\alpha]$

MT case: $\qquad \Theta = [\sin\alpha \; 0]$

i.e. in the BM case the longitudinal components are summed to give total signal, for MT the second component (taken to be the 'bound' pool) does not contribute. RF pulse transformation matrix $T_L$ is also defined differently for each case:

BM case: $\qquad T_L = \begin{bmatrix} \cos\alpha & 0 \\ 0 & \cos\alpha \end{bmatrix}$

MT case: $\qquad T_L = \begin{bmatrix} \cos\alpha & 0 \\ 0 & e^{-\overline{W(\omega_z)}\tau_{rf}} \end{bmatrix}$

Eq.[A1] is a generalization of the Ernst formula for a single pool system.

### Balanced SSFP

For bSSFP a coherent steady state between all components is considered. For the BM case written out for $M = \begin{bmatrix} M_+^a & M_-^a & M_z^a & M_+^b & M_-^b & M_z^b \end{bmatrix}^T$, the steady state signal is given by:

$$M_+^{ss} = \Theta \left\{\Delta - Te^{A^\dagger TR}\right\}^{-1} T\left\{e^{A^\dagger TR} - \mathbb{I}\right\} A^{\dagger^{-1}} C \qquad [A2]$$

where $A^\dagger$ is the full system matrix (Eq.[7]), $C = \begin{bmatrix} 0 & 0 & R_{1,a}M_0^a & 0 & 0 & R_{1,b}M_0^b \end{bmatrix}^T$, $T$ is the full 6x6 rotation matrix (Eq.[18]), $\Delta$ represents a rotation of 180° about the z-axis for both pools (accounts for the phase alternation in bSSFP), and $\Theta = [1\;0\;0\;1\;0\;0]$.

For the MT case the same expression is used except the rows and columns corresponding to $M_+^b$ and $M_-^b$ are deleted in all matrices, as are all coupling terms relating to $M_+^a$ and $M_-^a$. The system is a 4x4 matrix, the modified RF pulse matrix (Eq.[18]) is used, $C = \begin{bmatrix} 0 & 0 & R_{1,a}M_0^a & R_{1,b}M_0^b \end{bmatrix}^T$, and $\Theta = [1\;0\;0\;0]$.

**References**

1. Hennig J. Echoes—how to generate, recognize, use or avoid them in MR-imaging sequences. Part I: Fundamental and not so fundamental properties of spin echoes. Concepts Magn. Reson. 1991;3:125–143.

2. Hennig J. Echoes - how to generate, recognize, use or avoid them in MR-imaging sequences. Part II Echoes in Imaging Sequences. Concepts Magn. Reson. 1991;3:179–192.

3. Weigel M. Extended phase graphs: Dephasing, RF pulses, and echoes - pure and simple. J. Magn. Reson. Imaging 2015;41:266–295.

4. Hess AT, Robson MD. Hexagonal gradient scheme with RF spoiling improves spoiling performance for high-flip-angle fast gradient echo imaging. Magn. Reson. Med. doi: 10.1002/mrm.26213.

5. Nehrke K. On the steady-state properties of actual flip angle imaging (AFI). Magn. Reson. Med. 2009;61:84–92.

6. Hennig J, Scheffler K. Hyperechoes. Magn. Reson. Med. 2001;12:6–12.

7. Hennig J. Multiecho imaging sequences with low refocusing flip angles. J. Magn. Reson. 1988;407:397–407.

8. Hennig J, Weigel M, Scheffler K. Multiecho sequences with variable refocusing flip angles: optimization of signal behavior using smooth transitions between pseudo steady states (TRAPS). Magn. Reson. Med. 2003;49:527–35.

9. Weigel M, Hennig J. Contrast behavior and relaxation effects of conventional and hyperecho-turbo spin echo sequences at 1.5 and 3 T. Magn. Reson. Med. 2006;55:826–35.

10. Malik SJ, Padormo F, Price AN, Hajnal J V. Spatially resolved extended phase graphs: modeling and design of multipulse sequences with parallel transmission. Magn. Reson. Med. 2012;68:1481–94.

11. Sbrizzi A, Hoogduin H, Hajnal J V., Berg CAT Van Den, Luijten PR, Malik SJ, van den Berg CAT, Luijten PR, Malik SJ. Optimal control design of turbo spin-echo sequences with applications to parallel-transmit systems. Magn. Reson. Med. 2017;373:361–373.

12. Weigel M, Schwenk S, Kiselev VG, Scheffler K, Hennig J. Extended phase graphs with anisotropic diffusion. J. Magn. Reson. 2010;205:276–85.

13. Lankford CL, Dortch RD, Does MD. Fast $T_2$ mapping with multiple echo, caesar cipher acquisition and model-based reconstruction. Magn. Reson. Med. 2015;73:1065–1074.

14. Hamilton JI, Jiang Y, Chen Y, Ma D, Lo WC, Griswold M, Seiberlich N. MR fingerprinting for rapid quantification of myocardial T1, T2, and proton spin density. Magn. Reson. Med. doi: 10.1002/mrm.26216.

15. Cloos MA, Knoll F, Zhao T, Block K, Bruno M, Wiggins C, Sodickson D. Multiparamatric imaging with heterogenous radiofrequency fields. Nat. Commun. 2016;in press:1–10.

16. Prasloski T, Mädler B, Xiang QS, MacKay A, Jones C. Applications of stimulated echo correction to multicomponent T2 analysis. Magn. Reson. Med. 2012;67:1803–1814.

17. McConnell HM. Reaction Rates by Nuclear Magnetic Resonance. J. Chem. Phys. 1958;28:430–431.

18. Henkelman RM, Huang X, Xiang Q-SS, Stanisz GJ, Swanson SD, Bronskill MJ. Quantitative

interpretation of magnetization transfer. Magn Reson Med 1993;29:759–766.

19. Graham SJ, Henkelman RM. Understanding pulsed magnetization transfer. J. Magn. Reson. Imaging 1997;7:903–912.

20. Bieri O, Scheffler K. On the origin of apparent low tissue signals in balanced SSFP. Magn. Reson. Med. 2006;56:1067–74.

21. Ma D, Gulani V, Seiberlich N, Liu K, Sunshine JL, Duerk JL, Griswold MA. Magnetic resonance fingerprinting. Nature 2013;495:187–192.

22. Deoni SCL, Rutt BK, Arun T, Pierpaoli C, Jones DK. Gleaning Multicomponent T 1 and T 2 Information From Steady-State Imaging Data. 2008;1387:1372–1387.

23. Zaiss M, Bachert P. Exchange-dependent relaxation in the rotating frame for slow and intermediate exchange - modeling off-resonant spin-lock and chemical exchange saturation transfer. NMR Biomed. 2013;26:507–518.

24. Liu F, Block WF, Kijowski R, Samsonov A. Rapid multicomponent relaxometry in steady state with correction of magnetization transfer effects. Magn. Reson. Med. 2016;75:1423–1433.

25. Liu F, Velikina J V., Block WF, Kijowski R, Samsonov AA. Fast Realistic MRI Simulations Based on Generalized Multi-Pool Exchange Tissue Model. IEEE Trans. Med. Imaging 2017;36:527–537.

26. Morrison C, Mark Henkelman R. A Model for Magnetization Transfer in Tissues. Magn. Reson. Med. 1995;33:475–482.

27. Gloor M, Scheffler K, Bieri O. Quantitative magnetization transfer imaging using balanced SSFP. Magn. Reson. Med. 2008;60:691–700.

28. Ou X, Gochberg DF. MT effects and T1 quantification in single-slice spoiled gradient echo imaging. Magn. Reson. Med. 2008;59:835–845.

29. Liu F, Block WF, Kijowski R, Samsonov A. Rapid multicomponent relaxometry in steady state with correction of magnetization transfer effects. Magn. Reson. Med. doi: 10.1002/mrm.25672.

30. Malik SJ, Sbrizzi A, Hoogduin JM, Hajnal J V. Equivalence of EPG and Isochromat-Based Simulation of MR Signals. In: Proc ISMRM. ; 2016. p. 3196.

31. Pike GB. Pulsed Magnetization Transfer Contrast in Gradient Echo Imaging : A Two-Pool Analytic Description of Signal Response. Magn. Reson. Med. 1996;36:95–103.

32. Jiang Y, Ma D, Seiberlich N, Gulani V, Griswold MA. MR Fingerprinting Using Fast Imaging with Steady State Precession ( FISP ) with Spiral Readout. 2015;1631:1621–1631.

33. van Gelderen P, Jiang X, Duyn JH. Effects of magnetization transfer on T1 contrast in human brain white matter. Neuroimage 2016;128:85–95.

34. Yeung H, Aisen A. Magnetization Transfer Contrast with Periodic Pulsed Saturation. Radiology 1992;183:209–214.

35. MacKay a., Whittall K, Adler J, Li D, Paty D, Graeb D. In vivo visualization of myelin water in brain by magnetic resonance. Magn. Reson. Med. 1994;31:673–677.

36. Whittall KP, MacKay AL. Quantitative interpretation of NMR relaxation data. J. Magn. Reson. 1989;84:134–152.

37. Melki P., Mulkern R. Magnetization Transfer Effects in Multislice RARE Sequences. Magn. Reson. Med. 1992;195:189–195.

38. Weigel M, Helms G, Hennig J. Investigation and modeling of magnetization transfer effects in two-dimensional multislice turbo spin echo sequences with low constant or variable flip angles at 3 T. Magn. Reson. Med. 2010;63:230–234.

39. Yeung HN, Swanson SD. Transient decay of longitudinal magnetization in heterogeneous spin systems under selective saturation. J. Magn. Reson. 1992;99:466–479.

40. Alonso-Ortiz E, Levesque IR, Pike GB. MRI-based myelin water imaging: A technical review. Magn. Reson. Med. 2014;81:70–81.

41. Harkins KD, Dula AN, Does MD. Effect of Intercompartmental Water Exchange on the Apparent Myelin Water Fraction in Multiexponential T2 Measurements of Rat Spinal Cord. 2012;67:793–800.

42. Preibisch C, Deichmann R. Influence of RF spoiling on the stability and accuracy of T1 mapping based on spoiled FLASH with varying flip angles. Magn. Reson. Med. 2009;61:125–35.

43. Hamilton JI, Griswold M a, Seiberlich N. MR Fingerprinting with chemical exchange ( MRF-X ) to quantify subvoxel T1 and extracellular volume fraction. J. Cardiovasc. Magn. Reson. 2015;17:W35.

44. Stanisz GJ, Kecojevic A, Bronskill MJ, Henkelman RM. Characterizing White Matter With Magnetization Transfer and T2. Magn. Reson. Med. 1999;42:1128–1136.

45. Gloor M, Scheffler K, Bieri O. Nonbalanced SSFP-Based Quantitative Magnetization Transfer Imaging. Magn. Reson. Med. 2010;156:149–156.